\documentclass[conference]{IEEEtran}
\usepackage{graphicx} 
\usepackage[hidelinks]{hyperref}
\usepackage{todonotes}

\author{
\IEEEauthorblockN{Sarah Meriem Ourari}
\IEEEauthorblockA{
CrySyS Lab\\
Budapest Univ. of Technology and Economics, Hungary\\
sarah.ourari@crysys.hu\\
}
}

\title{Measuring the Security of the Evolving Software Supply Chain: a Research Agenda}

\begin{document}

\maketitle

\begin{abstract}
Software supply chain security has become increasingly critical due to the widespread reliance on third-party dependencies and the growing attack surface of modern software ecosystems. However, existing quantitative, measurement-based analysis and vulnerability management approaches remain largely fragmented and ecosystem-specific, limiting their ability to provide comparable risk assessments across environments.

This paper presents a structured research plan, starting with a Systematization of Knowledge (SoK) to synthesize the current state of research and identify key gaps, highlighting the limitations in dependency modeling and vulnerability propagation analysis, particularly in the treatment of transitive dependencies and their real-world exploitability. Based on these insights, we argue for a unified measurement perspective capable of consistently representing and analyzing the cross-ecosystem dependency structure.
We further identify emerging challenges introduced by AI-assisted software development, where coding LLMs are likely to contribute to new dependency patterns that are not captured by traditional Software Composition Analysis (SCA) tools. These shifts motivate a rethink of dependency modeling to account for evolving software-generation practices and their long-term structural impact on software security.
\end{abstract}

\begin{IEEEkeywords}
supply chain security, software security, vulnerability exposure
\end{IEEEkeywords}

\section{Introduction}

\begin{figure}[tb]
    \centering
\includegraphics[width=\columnwidth]{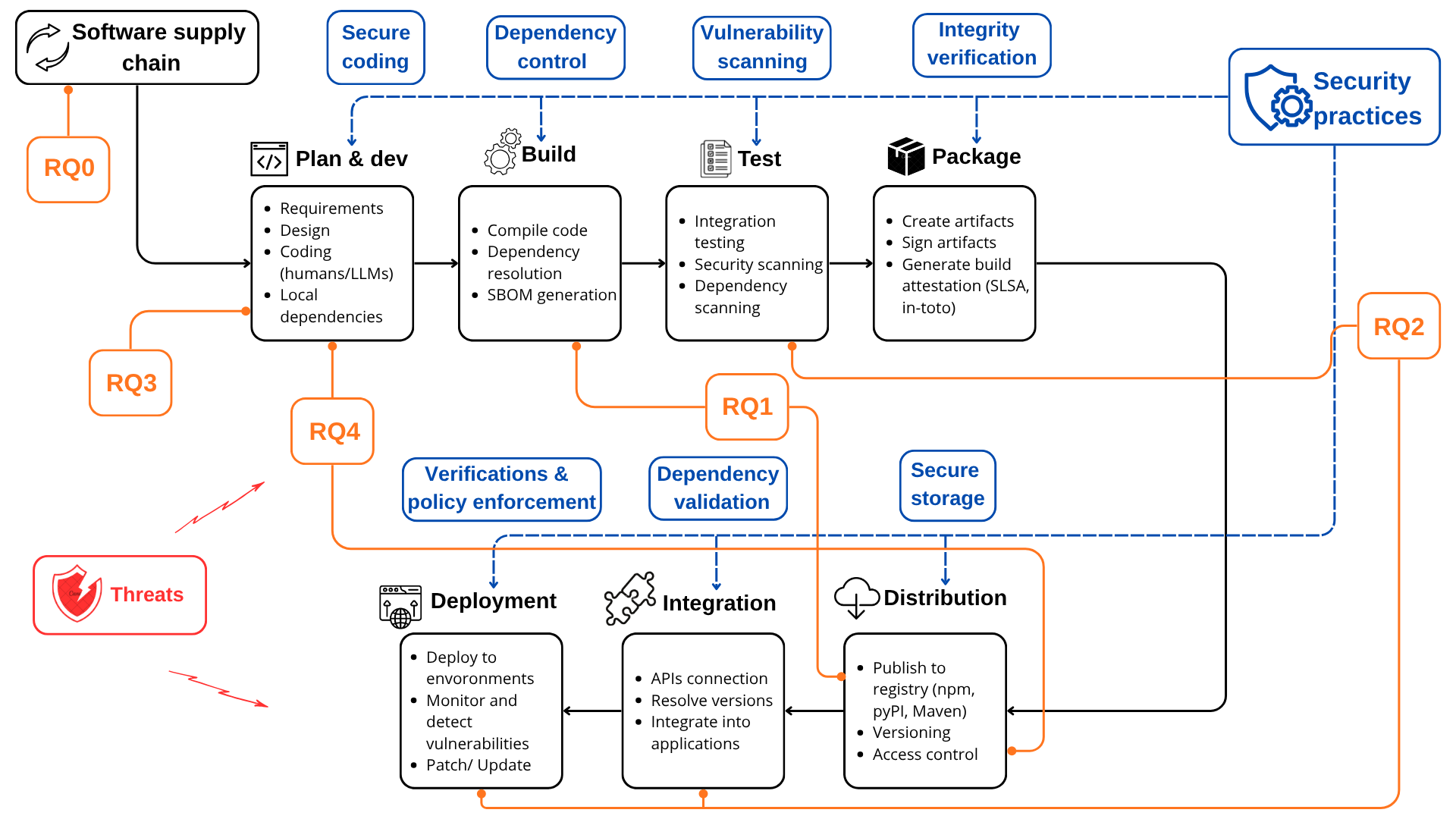}
    \caption{The Software Supply Chain Lifecycle and security mapping framework }
    \label{fig:SSCL}
\end{figure}

A software supply chain encompasses the entire end-to-end lifecycle of an application, from upstream development stages to downstream operational phases, as illustrated in Figure~\ref{fig:SSCL}. Securing this pipeline requires enforcing continuous security practices, such as secure coding, vulnerability scanning, and dependency validation, to protect both code generation and artifact delivery from malicious threats.

The importance of software supply chain security has grown significantly in recent years. Supply chain attacks have demonstrated that the compromise of a single dependency or maintainer can have widespread consequences across thousands of downstream applications. As a result, understanding and securing dependency relationships has become a central concern in both software engineering and cybersecurity research.

Despite this growing attention, several challenges remain unresolved. Security risks propagate not only through direct dependencies but also through transitive dependency chains, making vulnerability tracking highly complex. 
In addition, SCA tools often produce false positives, suffer from inaccurate dependency resolution, and fail to properly model real-world execution and reachability~\cite{zhao2023}.
Moreover, many existing approaches are fragmented and tailored to specific ecosystems, limiting their generalizability and making cross-ecosystem comparisons and risk assessments difficult.

In parallel, the increasing adoption of automated software development raises new questions for software supply chain security. Beyond traditional dependency management challenges, AI-generated code may introduce novel dependency relationships, influencing how developers select, reuse, and maintain software components. As a result, existing measurement and dependency modeling methodologies may need to be adapted/extended to accurately capture these emerging behaviors and their impact on software ecosystems.

In this context, this paper outlines future research aimed at developing a generalized framework for systematically measuring software security across multiple software ecosystems.
This study is driven by the need to bridge existing methodological gaps and explore how AI-assisted code generation affects ecosystem-level security analysis.
Based on the identified limitations and gaps, this work addresses the following research questions:

\begin{itemize}

\item \textbf{RQ0: What is the current state of the art in measuring and modeling software supply chain security across different software ecosystems? } 

\item \textbf{RQ1: How can a unified measurement model be designed to analyze dependency graphs and assess vulnerability exposure across diverse software ecosystems? }

\item \textbf{RQ2: How can source code-level reachability analysis reduce alert fatigue by identifying which vulnerabilities in transitive dependencies are actually executable?} 

\item \textbf{RQ3: To what extent does LLM-generated code introduce new forms of software dependencies, and how can the proposed measurement and dependency modeling framework be extended to capture them?}

\item \textbf{RQ4: How have dependency patterns evolved across software ecosystems following the adoption of LLM-assisted software development compared to the pre-LLM era?}
\end{itemize}


This paper is organized as follows. Section~\ref{sec:sok} presents a condensed systematic literature review on measurable software supply chain security. Section~\ref{sec:RQs} provides a synthesis of the main research gaps identified across multiple software ecosystems. Section~\ref{sec:methodology} outlines a future research agenda, 
while Section~\ref{sec:conclusion} concludes the paper.

\section{State of the Art\protect\footnote{a full-length Systematization of Knowledge (SoK) paper is under submission}}
\label{sec:sok}

A structured literature review methodology was followed to collect and analyze the relevant works in the field of software supply chain security. The articles were collected from scientific databases such as IEEE Xplore, ACM Digital Library, SpringerLink, Google Scholar, and Web of Science, with a focus on recent publications to ensure up-to-date findings. The selection focused on papers relevant to quantifiable software ecosystem security, dependency analysis, and vulnerability management, excluding non-scientific sources and low-quality papers.

\begin{figure}[t]
    \centering
    \includegraphics[width=\columnwidth]{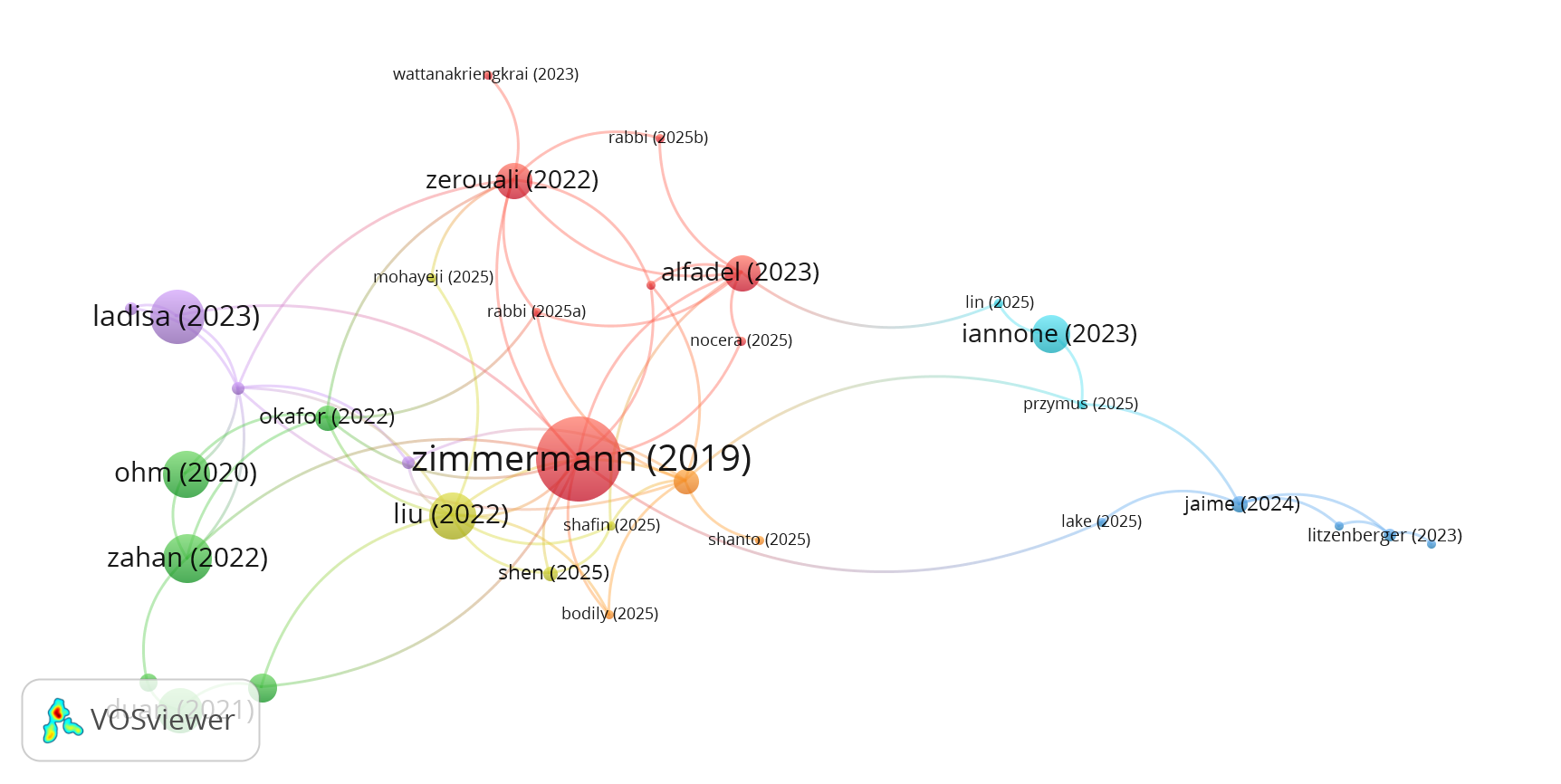}
    \caption{Citation network and clustering visualization using VOSviewer}
    \label{fig:VOSviewer}
\end{figure}

After filtering the articles, we constructed citation graphs using bibliometric analysis tools, e.g., VOSviewer and CiteNet Explorer, to visualize relationships among papers, identify influential works, and analyze citation patterns across studies. The resulting graph generated with VOSviewer is shown in Figure~\ref{fig:VOSviewer}.

Lastly, the selected literature was organized into thematic categories, including ecosystem-based and software metrics studies, SCA tools, vulnerability mitigation and patching techniques, and dependency graph structure and management. As part of the findings, several key results can be summarized as follows.

\subsection{Ecosystem risk dynamics and dependency metrics}
To map the global risk environment, researchers focused on quantifying how architectural choices and ecosystem management structures influence overall ecosystem risk. 
Various studies were conducted to identify major factors, assess the spread of vulnerabilities, and determine the real impact of dependency maintenance on ecosystem security. Zimmermann et al. \cite{zimmermann2019} investigated the influence of packages and maintainer accounts in compromising large parts of the ecosystem, focusing on direct dependencies, through reachability analysis and code vetting. Similarly, Zahan et al. \cite{zahan2022} empirically measured \emph{npm} weak links, enhancing previous metrics by introducing trust scores to capture vulnerability influence and by modeling collusion attacks to better characterize maintainer impact, highlighting both direct and transitive dependencies.

Taken together, these studies converge on two key structural patterns in software ecosystems: i) popular packages have a high-impact attack surface, as several libraries depend on them, and ii) a small number of highly influential maintainers affect a large portion of the ecosystem.

The study by Zahan et al. \cite{zahan2022} also reported several supply chain attack vectors, including install scripts, unmaintained packages, and expired domains, arising from maintainers being overloaded with a large number of packages.
To address these maintenance bottlenecks, Rahman et al.  \cite{rahman2025} emphasized the importance of dependency security by introducing two new metrics: \textit{MTTRdep\protect\footnote{MTTRdep (Mean Time To Remediate dependencies): Average time required to remediate vulnerable dependencies.}} and \textit{MTTUdep.\protect\footnote{MTTUdep (Mean Time To Update dependencies): Average time required to update dependencies.}}
Their quantitative results indicate that the effectiveness of dependency updates reflects how well a package addresses vulnerabilities, suggesting that improving update practices can significantly strengthen overall security.

Beyond structural dependency risks, researchers have also investigated how vulnerabilities evolve and are managed across different software ecosystems. Alfadel et al.~\cite{alfadel2023} studied the \emph{PyPI} ecosystem in comparison with \emph{npm} to understand its specific characteristics, while also developing DEPHEALTH, a tool designed to detect and report vulnerabilities. On the other hand, Zerouali et al.~\cite{zerouali2022} examined both \emph{npm} and \emph{RubyGems} ecosystems. Synthesizing findings across these environments, both studies reported that most discovered vulnerabilities are of medium or high severity, with Cross-Site Scripting (XSS) being the most common vulnerability type. Another critical cross-ecosystem finding concerns the differences in disclosure speed. For instance, \emph{npm} appears to have faster patching of identified vulnerabilities, but may still leave some issues unresolved for longer periods due to unknown real discovery dates and a lack of CVE identifiers. In contrast, \emph{PyPI} tends to have longer update delays, as vulnerabilities are disclosed more quickly, reducing the remediation window compared to \emph{npm}.
Despite these operational differences, both studies emphasize the role of indirect dependencies in increasing exposure, highlighting the importance of timely updates.

Finally, focusing on the \emph{Maven} ecosystem, Rabbi et al.~\cite{rabbi2025} and Nachuma et al.~\cite{nachuma2025} analyzed vulnerability evolution and dependency management processes using the Goblin framework~\cite{jaime2024} and the CWE database. These studies concluded that most vulnerabilities stem from recurring software weaknesses, particularly issues in input validation and access control, unencrypted data handling, and authentication failures. In addition, the patching process was found to be relatively slow, highlighting the need for faster vulnerability discovery, reporting, and documentation.

\subsection{Software Composition Analysis and reachability}
In order to analyze software dependencies and detect these security vulnerabilities, SCA tools are commonly used in research. Studies revealed several limitations in their ability to handle the complexity of software ecosystems, as most of them focus only on direct dependencies while ignoring dependency resolution issues, leading to false vulnerability warnings and reduced accuracy. Zhao et al. \cite{zhao2023} investigated the effectiveness of SCA tools in the Java \emph{Maven} ecosystem by proposing a new evaluation model that considers execution environments, dependency and vulnerability detection mechanisms, and dependency selection. Results show difficulties in handling cross-project information, such as parent configurations and third-party component management, leading to missing inherited packages or incorrect versions in pre-built scans. In addition, there is no distinction between the types of external libraries. As a result, system dependencies are integrated into the software, increasing false positives. The study also revealed that vulnerability detection performance heavily relies on dependency analysis. Therefore, research confirms the need to use the developed Scan Scope Maven (SSM) concept, which reduces both false positives and false negatives by defining the required dependencies to include during project analysis. Furthermore, researchers highlight the need for SCA tools to extend detection beyond package managers to include external references and copy-pasted code~\cite{zhao2023}, which represent a large part of the \emph{Maven} ecosystem, and to choose the tools based on the use case since the execution environment considerably influences the results.
These limitations have also been observed across other SCA tools and ecosystems, suggesting that the issue is not tool-specific but fundamentally related to dependency resolution and reachability modeling.

Building on those observations, subsequent work has examined SCA tools in real-world settings, focusing on how vulnerability detection outcomes influence developers' decision-making. Nocera et al. \cite{nocera2025}, through their cohort study, confirm previous findings regarding the suitability of the OWASP Dependency-Check tool for built projects, with results showing a considerable reduction in the number and severity of vulnerabilities, especially high-severity and informational vulnerabilities. Similarly, Mohayeji et al. \cite{mohayeji2025} extended this research by investigating the practical usefulness of Dependabot through an analysis of developers' reactions to its security updates in the \emph{npm} ecosystem. Results show that in more than half of the cases, developers directly apply the fixes suggested by Dependabot without observing major issues after integration. In addition, the study revealed that automated fixes are more frequent than manual ones, except for major version upgrades due to compatibility constraints. Additional improvements suggest incorporating vulnerability reachability analysis, as some detected weaknesses may not be reached and therefore may not represent relevant threats. This could help reduce unnecessary alerts and decrease false positives.

To improve the precision of vulnerability detection, recent research has shifted toward reachability analysis and graph-based dependency modeling techniques. Wu et al.~\cite{wu2023} explored source code-level analysis to better detect actual risks. The study shows that most identified vulnerable functions do not pose a real danger in practice, as only a small percentage is reachable, and many vulnerabilities are either deeply embedded in the dependency tree or require complex conditions to be exploited. Therefore, SCA tools frequently raise alerts even when there is no real danger, overloading developers with unnecessary updates. An additional factor contributing to the inaccuracy of SCA tools is the presence of ecosystem-specific characteristics. Liu et al.~\cite{liu2022} developed a solution to overcome limitations of \emph{npm} rules by constructing a knowledge graph linking packages to their known vulnerabilities, along with an algorithm to accurately reconstruct the true dependency graph, achieving more precise results.

\subsection{Vulnerability remediation and lifecycle principles}
To address challenges in vulnerability management,
Iannone et al. \cite{iannone2023} emphasized the importance of extending existing analysis tools by incorporating development timelines, developer actions, and contextual changes into vulnerability assessment. Similarly, Okafor et al. \cite{okafor2022} introduced three key security principles combining dependency transparency, component validity, and the separation of development stages to support ecosystem-wide analysis.

Building on these conceptual foundations, several studies have proposed automated approaches to improve vulnerability remediation in practice. Zhang et al.~\cite{zhang2023} introduced Ranger, a tool that automatically restores version ranges while preserving compatibility with existing code bases, and thus, reducing the risk of breaking changes and limiting the persistence of security flaws. 
However, accurately linking vulnerabilities to their corresponding fixes remains a challenging problem. Li et al. \cite{li2024} address this issue with PatchFinder, a system that identifies relevant patch commits by leveraging semantic similarity and contextual code understanding, thereby improving the reliability of the results.

\subsection{Dependency graph construction and mining frameworks}
Dependency graphs, as one of the primary artifacts used in software ecosystem analysis, have received considerable attention due to the growing complexity of software supply chains and the increasing number of attacks. To address the lack of standardization in dependency graph construction, Litzenberger et al.~\cite{litzenberger2023} proposed the Dependency Graph Mining Framework (DGMF) as a unified solution for building dependency graphs from different software repositories. The results demonstrated improved execution efficiency through reduced runtime and simplified adaptation to heterogeneous repositories. However, a trade-off still occurs during dependency resolution, which comes at the cost of increased computational efforts.

Building on the need for more accurate dependency representations, Jaime et al.~\cite{jaime2024} developed the Goblin framework 
to support historical analysis. The framework reconstructs dependency graphs while incorporating temporal information, enabling more precise analysis and on-demand computation of ecosystem metrics, which improves dependency quality. Their findings revealed several factors that can influence research outcomes, including differences in dataset size and structure, dependency updates and management, and the high computational costs associated with metric processing across transitive dependencies.

Beyond the technical challenges of dependency graph construction and analysis, researchers have also identified inefficiencies in the research process. Many studies repeatedly rebuild the same basic custom tools to collect and process ecosystem data, leading to duplicated effort, limited reusability, and a large number of abandoned or unmaintained research artifacts. 
To address these challenges, Dusing et al.~\cite{dusing2025} proposed the MARIN framework, which automates repetitive technical tasks in \emph{Maven} ecosystem analysis, while improving performance through parallel processing. Similar efforts have also been undertaken in other software ecosystems. Filgueira et al.~\cite{filgueira2022} developed the static analysis framework \textit{inspect4py} for Python, which examines source code structure and extracts relevant software artifacts to support software engineering and ecosystem analysis tasks.

These findings reflect the current state of problems and potential solutions aimed at improving measurable aspects of software supply chain security; however, a lack of generalizability is evident. Most proposed solutions are ecosystem-specific and rely on particular characteristics, making comparison across studies difficult. This highlights the need for more comprehensive metrics for mitigation and management processes.

\section{Literature synthesis and research questions}\label{sec:RQs}

Based on previous studies, several important findings have contributed to a better understanding of software security chains. At the same time, the adoption of Large Language Models (LLMs) has introduced new challenges and transformations in software development and security analysis, highlighting the need for further investigations into their impact.

Although existing solutions make valuable contributions to vulnerability management and software security, most proposed approaches and metrics remain ecosystem-specific and rely on specific characteristics. This lack of generalization limits the development of unified mitigation and management strategies by centralizing research efforts on analyzing individual environments in isolation. As a result, there is no guarantee that approaches effective in one ecosystem can be directly transferred to another. Therefore, the evaluation of such solutions remains constrained to the datasets and the conditions of a single ecosystem. Furthermore, this fragmented perspective limits the exploration of structural similarities and differences across ecosystems, potentially leading to an incomplete understanding of the global problem space. 
Based on these observations, we formulate~\textbf{RQ1: How can a unified measurement model be designed to analyze dependency graphs and assess vulnerability exposure across diverse software ecosystems?}

In addition, most studies focus heavily on direct dependencies while oversimplifying the impact of transitive and third-party packages, which have a significant influence on ecosystem security. A common operational limitation is the difficulty of correctly resolving dependency versions, often driven by developers' fear of introducing breaking updates. 
Furthermore, not all vulnerabilities present in a dependency tree are necessarily reachable during program execution. As a result, developers are overwhelmed with excessive updates generated by SCA tools, which often miss certain packages or introduce unrelated dependencies into the analysis~\cite{zhao2023}, reducing the credibility of the findings due to inaccuracies in the extracted graphs.
This highlights an urgent need for more fine-grained analysis that can distinguish which updates are truly necessary and have a meaningful impact on ecosystem security. 
Such improvement can be achieved through source code-level reachability analysis to identify exploitable vulnerabilities and support more relevant dependency updates. Existing tools such as \textit{CodeQL\protect\footnote{static analysis engine developed by Github for querying source code as a database.}} illustrate how source-code analysis can be used to determine whether vulnerable code paths are actually reachable during execution. In addition, this perspective should also account for cross-project reuse and code cloning, as they represent a significant portion of software ecosystems.
Based on these observations, we formulate~\textbf{RQ2: How can source-code-level reachability analysis reduce alert fatigue by identifying which vulnerabilities in transitive dependencies are actually executable?}

While optimizing traditional dependency mapping represents a crucial foundational step, the modern software supply chain is facing an unpredictable paradigm shift due to the rapid integration of artificial intelligence in software development.
The widespread adoption of LLMs for automated code generation has changed how applications are written, directly impacting how dependencies are introduced. 
Package recommendations generated by AI models may introduce distinct forms of software dependencies and hidden security risks at the source level. It is therefore critical to investigate how the foundational measurement and graph modeling frameworks must evolve to effectively capture this new class of supply chain elements. This leads to the next research question, \textbf{RQ3: To what extent does LLM-generated code introduce new forms of software dependencies, and how can the proposed measurement and dependency modeling framework be extended to capture them?}
In parallel, as AI-based tools become integrated into development workflows, both the volume of generated code and the practices surrounding code reuse are evolving. Understanding the implications of this shift requires comparing dependency structures before and after the widespread use of LLMs to reveal how dependency relationships, depth, and versioning practices have evolved over time. Consequently, this study seeks to quantify this historical transition by tackling~\textbf{RQ4: How have dependency patterns evolved across software ecosystems following the adoption of LLM-assisted software development, compared to the pre-LLM era?}

\section{Proposed research methodology}\label{sec:methodology}

\begin{figure}[tb]
    \centering
    \includegraphics[width=\columnwidth]{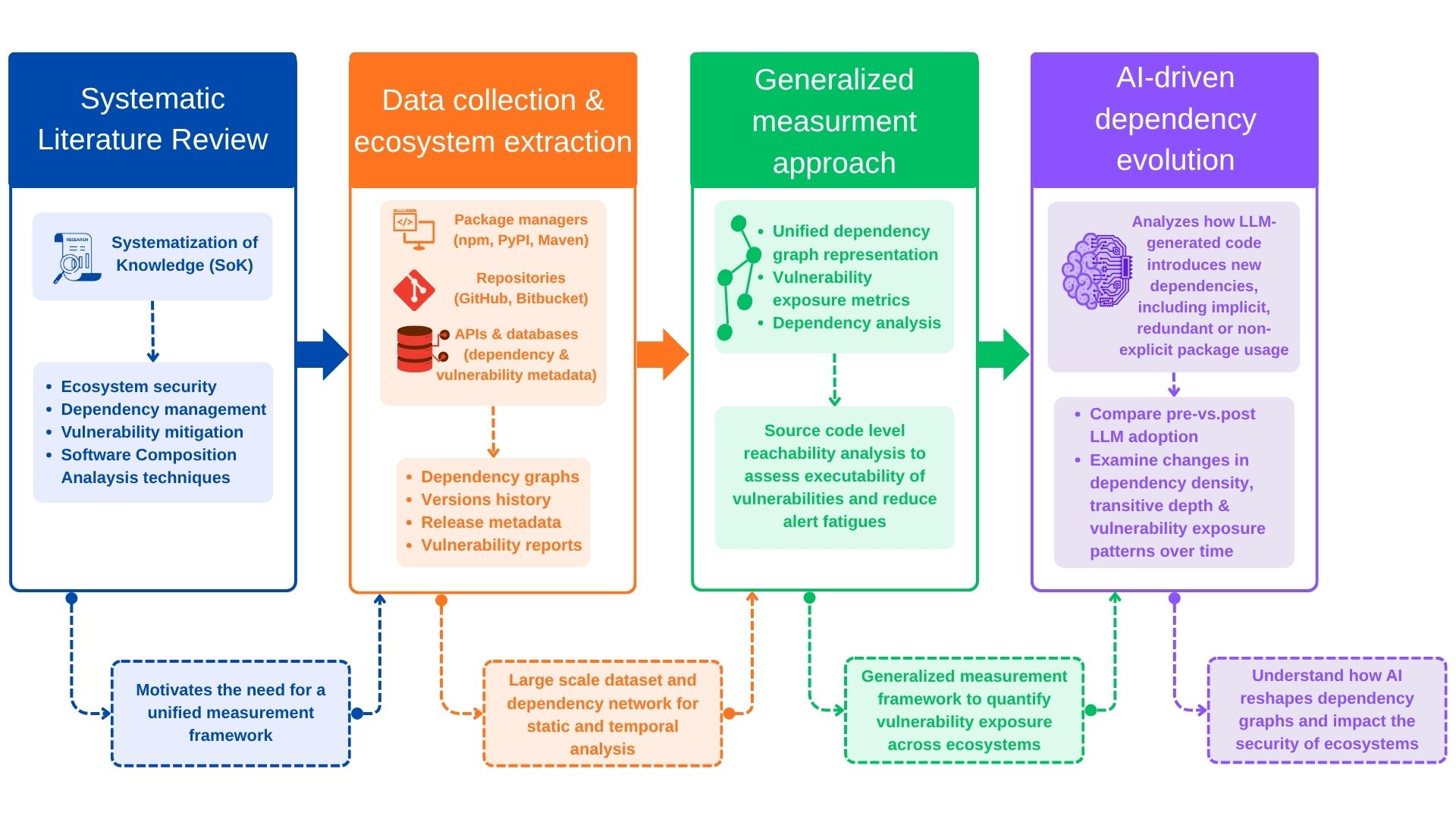}
    \caption{Main phases of the proposed research methodology}
    \label{fig:methodology}
\end{figure}

The adopted methodology structuring this research can be summarized in three main stages: literature review and state of the art, the design of a generalized measurement approach, and the investigation of its applicability for improving software supply chain security. The overall workflow of the proposed methodology is illustrated in Figure~\ref{fig:methodology}.

\subsection{Systematic literature review}

The first phase consists of an analysis of existing studies on software system security, dependency management, and vulnerability mitigation, culminating in the writing of a SoK paper to understand the current state of research and analyze existing tools and metrics. From this literature overview, key limitations and research gaps are identified to highlight shortcomings in current approaches, leading to the formulation of the main research questions, which focus on the need for a more generalizable framework for ecosystem-level analysis.

\subsection{Data collection and ecosystem extraction}

This phase focuses on building a large-scale dataset of software ecosystems to support empirical analysis. Data is collected from multiple sources, including package managers (e.g., \emph{npm}, \emph{PyPI}, \emph{Maven}), source code repositories such us \textit{GitHub} and \textit{Bitbucket}, and dependency metadata APIs.
The collected data includes dependency graphs, version histories, release metadata, and vulnerability reports.
The extracted datasets are then used to construct heterogeneous dependency networks across ecosystems, enabling both static and temporal analysis of software supply chains.

\subsection{Design of a generalized measurement approach}

The core contribution of this research lies in exploring the feasibility of a unified measurement model that captures and enables the evaluation of dependency-related risks and ecosystem-level security exposure across different software environments.

First, a unified dependency graph representation is developed to capture direct and transitive relationships across heterogeneous ecosystems. Based on this representation, a measurement framework is defined to quantify vulnerability exposure, dependency risk propagation, and ecosystem-level security impact.

To refine the vulnerability assessment, a source-code-level reachability analysis is integrated to determine whether vulnerabilities in transitive dependencies are actually executable in practice. This mechanism aims to reduce alert fatigue by distinguishing between theoretical exposure and real exploitability.

\subsection{LLM-driven code generation and ecosystem evolution analysis}

The last phase focuses on the impact of LLM-assisted software development code on dependency ecosystems. First, the study analyzes how LLM-generated code introduces new dependencies, including implicit, redundant, or non-explicit package usages. The previously defined measurement and modeling framework is extended to capture these newly emerging dependency patterns, potentially based on probabilistic graphical models~\cite{pgm}.
Second, a temporal analysis is conducted to compare dependency structures before and after the widespread adoption of LLM-assisted development tools. This includes examining changes in dependency density, transitive depth, and vulnerability exposure patterns across ecosystems over time.

\section{Conclusion}\label{sec:conclusion}

This paper presents a 
research plan based on the current state of the literature, derived from a SoK paper on quantifiable software supply chain security, covering ecosystem security, dependency management, vulnerability mitigation techniques, and SCA tools.

The review highlights the absence of a generalized measurement framework for consistent ecosystem-level security evaluation, motivating a research direction focused on the design of a unified software supply chain security measurement framework across multiple ecosystems, as well as an investigation into how artificial intelligence and LLMs reshape software dependency structures.

\newpage

\nocite{*}
\bibliographystyle{IEEEtran}
\bibliography{references}

@article{alfadel2023,
  author  = {Alfadel, M. and Costa, D. E.},
  title   = {Empirical Analysis of Security Vulnerabilities in Python Packages},
  journal = {Empirical Software Engineering},
  volume  = {28},
  number  = {1},
  pages   = {1--34},
  year    = {2023},
  doi     = {10.1007/s10664-022-10278-4},
  url     = {https://doi.org/10.1007/s10664-022-10278-4}
}

@inproceedings{dusing2025,
  author    = {D{\"u}sing, J. and Chiaramonte, J. and Hermann, B.},
  title     = {MARIN: A Research-Centric Interface for Querying Software Artifacts on Maven Repositories},
  booktitle = {2025 IEEE/ACM 22nd International Conference on Mining Software Repositories (MSR)},
  pages     = {591--595},
  year      = {2025},
  doi       = {10.1109/MSR66628.2025.00093},
  url       = {https://doi.org/10.1109/MSR66628.2025.00093}
}

@inproceedings{filgueira2022,
  author    = {Filgueira, R. and Garijo, D.},
  title     = {Inspect4py: A Knowledge Extraction Framework for Python Code Repositories},
  booktitle = {Proceedings of the 19th International Conference on Mining Software Repositories},
  pages     = {232--236},
  year      = {2022},
  doi       = {10.1145/3524842.3528497},
  url       = {https://doi.org/10.1145/3524842.3528497}
}

@article{iannone2023,
  author  = {Iannone, E. and Guadagni, R. and Ferrucci, F. and De Lucia, A. and Palomba, F.},
  title   = {The Secret Life of Software Vulnerabilities: A Large-Scale Empirical Study},
  journal = {IEEE Transactions on Software Engineering},
  volume  = {49},
  number  = {1},
  pages   = {44--63},
  year    = {2023},
  doi     = {10.1109/TSE.2022.3140868},
  url     = {https://doi.org/10.1109/TSE.2022.3140868}
}

@inproceedings{jaime2024,
  author    = {Jaime, D. and Haddad, J. E. and Poizat, P.},
  title     = {Goblin: A Framework for Enriching and Querying the Maven Central Dependency Graph},
  booktitle = {Proceedings of the 21st International Conference on Mining Software Repositories},
  pages     = {37--41},
  year      = {2024},
  doi       = {10.1145/3643991.3644879},
  url       = {https://doi.org/10.1145/3643991.3644879}
}

@inproceedings{li2024,
  author    = {Li, K. and Zhang, J. and Chen, S. and Liu, H. and Liu, Y. and Chen, Y.},
  title     = {PatchFinder: A Two-Phase Approach to Security Patch Tracing for Disclosed Vulnerabilities in Open-Source Software},
  booktitle = {Proceedings of the 33rd ACM SIGSOFT International Symposium on Software Testing and Analysis},
  pages     = {590--602},
  year      = {2024},
  doi       = {10.1145/3650212.3680305},
  url       = {https://doi.org/10.1145/3650212.3680305}
}

@inproceedings{litzenberger2023,
  author    = {Litzenberger, T. and D{\"u}sing, J. and Hermann, B.},
  title     = {DGMF: Fast Generation of Comparable, Updatable Dependency Graphs for Software Repositories},
  booktitle = {2023 IEEE/ACM 20th International Conference on Mining Software Repositories (MSR)},
  pages     = {115--119},
  year      = {2023},
  doi       = {10.1109/MSR59073.2023.00028},
  url       = {https://doi.org/10.1109/MSR59073.2023.00028}
}

@inproceedings{liu2022,
  author    = {Liu, C. and Chen, S. and Fan, L. and Chen, B. and Liu, Y. and Peng, X.},
  title     = {Demystifying the Vulnerability Propagation and Its Evolution via Dependency Trees in the NPM Ecosystem},
  booktitle = {Proceedings of the 44th International Conference on Software Engineering},
  pages     = {672--684},
  year      = {2022},
  doi       = {10.1145/3510003.3510142},
  url       = {https://doi.org/10.1145/3510003.3510142}
}

@article{mohayeji2025,
  author  = {Mohayeji, H. and Agaronian, A. and Constantinou, E. and Zannone, N. and Serebrenik, A.},
  title   = {Securing Dependencies: A Comprehensive Study of Dependabot’s Impact on Vulnerability Mitigation},
  journal = {Empirical Software Engineering},
  volume  = {30},
  number  = {3},
  pages   = {89},
  year    = {2025},
  doi     = {10.1007/s10664-025-10638-w},
  url     = {https://doi.org/10.1007/s10664-025-10638-w}
}

@inproceedings{nachuma2025,
  author    = {Nachuma, C. and Hossan, M. M. and Turzo, A. K. and Zibran, M. F.},
  title     = {Decoding Dependency Risks: A Quantitative Study of Vulnerabilities in the Maven Ecosystem},
  booktitle = {2025 IEEE/ACM 22nd International Conference on Mining Software Repositories (MSR)},
  pages     = {270--274},
  year      = {2025},
  doi       = {10.1109/MSR66628.2025.00048},
  url       = {https://doi.org/10.1109/MSR66628.2025.00048}
}

@inproceedings{nocera2025,
  author    = {Nocera, S. and Vegas, S. and Scanniello, G. and Juristo, N.},
  title     = {Software Composition Analysis and Supply Chain Security in Apache Projects: An Empirical Study},
  booktitle = {2025 IEEE/ACM 22nd International Conference on Mining Software Repositories (MSR)},
  pages     = {103--115},
  year      = {2025},
  doi       = {10.1109/MSR66628.2025.00027},
  url       = {https://doi.org/10.1109/MSR66628.2025.00027}
}

@inproceedings{okafor2022,
  author    = {Okafor, C. and Schorlemmer, T. R. and Torres-Arias, S. and Davis, J. C.},
  title     = {SoK: Analysis of Software Supply Chain Security by Establishing Secure Design Properties},
  booktitle = {Proceedings of the 2022 ACM Workshop on Software Supply Chain Offensive Research and Ecosystem Defenses},
  pages     = {15--24},
  year      = {2022},
  doi       = {10.1145/3560835.3564556},
  url       = {https://doi.org/10.1145/3560835.3564556}
}

@inproceedings{rabbi2025,
  author    = {Rabbi, M. F. and Paul, R. and Champa, A. I. and Zibran, M. F.},
  title     = {Understanding Software Vulnerabilities in the Maven Ecosystem: Patterns, Timelines, and Risks},
  booktitle = {2025 IEEE/ACM 22nd International Conference on Mining Software Repositories (MSR)},
  pages     = {290--294},
  year      = {2025},
  doi       = {10.1109/MSR66628.2025.00052},
  url       = {https://doi.org/10.1109/MSR66628.2025.00052}
}

@article{rahman2025,
  author  = {Rahman, I. and Paramitha, R. and Enck, W. and Williams, L.},
  title   = {How Quickly Do Development Teams Update Their Vulnerable Dependencies?},
  journal = {arXiv preprint arXiv:2403.17382},
  year    = {2025},
  doi     = {10.48550/arXiv.2403.17382},
  url     = {https://doi.org/10.48550/arXiv.2403.17382}
}

@inproceedings{wu2023,
  author    = {Wu, Y. and Yu, Z. and Wen, M. and Li, Q. and Zou, D. and Jin, H.},
  title     = {Understanding the Threats of Upstream Vulnerabilities to Downstream Projects in the Maven Ecosystem},
  booktitle = {2023 IEEE/ACM 45th International Conference on Software Engineering (ICSE)},
  pages     = {1046--1058},
  year      = {2023},
  doi       = {10.1109/ICSE48619.2023.00095},
  url       = {https://doi.org/10.1109/ICSE48619.2023.00095}
}

@inproceedings{zahan2022,
  author    = {Zahan, N. and Zimmermann, T. and Godefroid, P. and Murphy, B. and Maddila, C. and Williams, L.},
  title     = {What are Weak Links in the npm Supply Chain?},
  booktitle = {Proceedings of the 44th International Conference on Software Engineering: Software Engineering in Practice},
  pages     = {331--340},
  year      = {2022},
  doi       = {10.1145/3510457.3513044},
  url       = {https://doi.org/10.1145/3510457.3513044}
}

@article{zerouali2022,
  author  = {Zerouali, A. and Mens, T. and Decan, A. and Roover, C. D.},
  title   = {On the Impact of Security Vulnerabilities in the npm and RubyGems Dependency Networks},
  journal = {Empirical Software Engineering},
  volume  = {27},
  number  = {5},
  year    = {2022},
  doi     = {10.1007/s10664-022-10154-1},
  url     = {https://doi.org/10.1007/s10664-022-10154-1}
}

@inproceedings{zhang2023,
  author    = {Zhang, L. and Liu, C. and Chen, S. and Xu, Z. and Fan, L. and Zhao, L. and Zhang, Y. and Liu, Y.},
  title     = {Mitigating Persistence of Open-Source Vulnerabilities in Maven Ecosystem},
  booktitle = {2023 38th IEEE/ACM International Conference on Automated Software Engineering (ASE)},
  pages     = {191--203},
  year      = {2023},
  doi       = {10.1109/ASE56229.2023.00058},
  url       = {https://doi.org/10.1109/ASE56229.2023.00058}
}

@inproceedings{zhao2023,
  author    = {Zhao, L. and Chen, S. and Xu, Z. and Liu, C. and Zhang, L. and Wu, J. and Sun, J. and Liu, Y.},
  title     = {Software Composition Analysis for Vulnerability Detection: An Empirical Study on Java Projects},
  booktitle = {Proceedings of the 31st ACM Joint European Software Engineering Conference and Symposium on the Foundations of Software Engineering},
  pages     = {960--972},
  year      = {2023},
  doi       = {10.1145/3611643.3616299},
  url       = {https://doi.org/10.1145/3611643.3616299}
}

@article{zimmermann2019,
  author  = {Zimmermann, M. and Staicu, C.-A. and Tenny, C. and Pradel, M.},
  title   = {Small World with High Risks: A Study of Security Threats in the npm Ecosystem},
  journal = {arXiv preprint arXiv:1902.09217},
  year    = {2019},
  doi     = {10.48550/arXiv.1902.09217},
  url     = {https://arxiv.org/abs/1902.09217}
}

@book{pgm,
  title={Probabilistic graphical models: principles and techniques},
  author={Koller, Daphne and Friedman, Nir},
  year={2009},
  publisher={MIT press}
}
\end{document}